\documentclass[journal]{IEEEtran}

\usepackage{cite}                       
\usepackage[inline]{enumitem}           
\usepackage[bookmarks=false]{hyperref}  
\usepackage[utf8]{inputenc}             
\usepackage{graphicx}                   
\usepackage{cite}                       
\usepackage[scaled=.92]{helvet}         
\usepackage{booktabs}                   
\usepackage{calc}                       
\usepackage[table,xcdraw]{xcolor}
\usepackage[normalem]{ulem}
\usepackage{multirow}                   
\usepackage{balance}                    
\usepackage{url}                        
\usepackage{breakurl}                   
\usepackage[cmex10]{amsmath}            
\usepackage{amsmath}
\usepackage{epstopdf}                   
\usepackage{svg}                        
\usepackage[inline]{enumitem}           
\usepackage{soul}                       
\usepackage{cuted}                      
\usepackage{ragged2e}                   
\usepackage[table,xcdraw]{xcolor}       
\usepackage{gensymb}                    
\usepackage[font=footnotesize]{caption} 
\usepackage{subcaption}                 
\usepackage{float}                      
\usepackage{colortbl}                   
\usepackage{pdflscape}                  
\usepackage{caption}                    
\usepackage{array}
\usepackage{multirow, multicol, makecell, booktabs}
\usepackage{authblk}

\definecolor{skin}{rgb}{1,0.88,0.74}    

\usepackage{url}
\usepackage{tabularx}

\begin{document}



\title{Security for Emerging Miniaturized Wireless Biomedical Devices: Threat Modeling with Application to Case Studies}

\author{Vladimir~Vakhter$^1$, Betul~Soysal, Patrick~Schaumont$^1$, and Ulkuhan~Guler$^1$\vspace{-4mm}
\thanks{$^1$Department of Electrical and Computer Engineering, Worcester Polytechnic Institute, Worcester, USA
        {\tt\scriptsize \{vvvakhter;~pschaumont;~uguler\}@wpi.edu, betuls@ieee.org}.
        }
}

\maketitle

\begin{abstract}
\thinspace The landscape of miniaturized wireless biomedical devices (MWBDs) is rapidly expanding as proactive mobile healthcare proliferates. MWBDs are diverse and include various injectable, ingestible, implantable, and wearable devices. While the growth of MWBDs increases the flexibility of medical services, the adoption of these technologies brings privacy and security risks for their users. MWBDs can operate with sensitive, private information and affect patients through the use of stimulation and drug delivery. Therefore, these devices require trust and need to be secure. Embedding protective mechanisms into MWBDs is challenging because they are restricted in size, power budget, as well as processing and storage capabilities. Nevertheless, MWBDs need to be at least minimally securable in the face of evolving threats. The main intent of this work is to make the primary stakeholders of MWBDs aware of associated risks and to help the architects and the manufacturers of MWBDs protect their emerging designs in a repeatable and structured manner. Making MWBDs securable begins with performing threat modeling. This paper introduces a domain-specific qualitative-quantitative threat model dedicated to MWBDs. The proposed model is then applied to representative case studies from each category of MWBDs. \end{abstract}

\IEEEpeerreviewmaketitle

\section{Introduction}
\label{sec:introduction}

\IEEEPARstart{R}{}ising interest in remote health monitoring and treatment stimulates an increase in the variety and the volume of miniaturized wireless biomedical devices (MWBDs)~\cite{bettinger2018advances, koydemir2018wearable, dunn2018wearables, fagan_foundational_2020}. The need for these smart-and-connected health technologies is foreseen to remain rising globally as their share in the reduction of healthcare costs grows~\cite{insider2019hiot}. Being convenient, low-cost, and easy-access, MWBDs support a transition from the traditional reactive medicine to the proactive, personalized precision healthcare model~\cite{kiourti2017review, wg_interagency_2018}. We started to witness more interactive communication between patients and healthcare providers with the help of remote monitoring devices, including MWBDs, especially after the global crisis of the public health system caused by the COVID-19 pandemic~\cite{it2020covid, 2020guler}. This trend is expected to create even more demand for MWBDs among patients and healthcare providers~\cite{deloitte2020hcare, med2020telemed}.

While novel designs are regularly presented, more applications and use scenarios are envisioned for MWBDs. Therefore, they are still considered emerging devices. MWBDs may be divided into four main categories:
\begin{enumerate*} [label=(\arabic*)]
    \item injectables, injected underneath the human tissue;
    \item implantables, implanted into the human body during a surgery;
    \item ingestibles, ingested by the patient in the form of a regular pill; and
    \item wearables, worn on the human body.
\end{enumerate*}

MWBDs are capable of collecting and transmitting sensitive, private information, like bio-electrical activity~\cite{song2019soc,li2020wearable} and vital signs~\cite{jiang2019biomote, mimee2018ingestible}, and affecting of the human body through stimulation~\cite{charthad2018stimulation, johnson2018stimdust} and drug delivery~\cite{yan2019battery, guo2019novel}. Therefore, while being convenient, they produce privacy and security risks for their users~\cite{selimis2020security, sun2018security, alexander2019implanted, kotz2016privacy}. Traditionally, designers and manufacturers of MWBDs tend to prioritize functionality and user experience over security~\cite{sun2018security, alexander2019implanted, kotz2016privacy}. As a result, protection mechanisms are missing at the architectural level for most of MWBDs~\cite{li2020wearable, jiang2019biomote, mimee2018ingestible, charthad2018stimulation, johnson2018stimdust, yan2019battery, guo2019novel}. Multiple attacks for the misuse of sensitive medical information and the malfunctioning of MWBDs may be implemented~\cite{camara2015security}. Professionals are now tasked with defeating well-funded attacks that, in some cases, can cause immediate physical harm for the user~\cite{megas_internet_2017, wired2018pacemaker, cnbc2018pacemaker, healthitsec2018pacemaker}. Attackers may also be attracted by the assets belonging to other primary stakeholders of MWBDs, as described in Table~\ref{tab:attackers_motivation}.

\begin{table}[h!]
\renewcommand{\arraystretch}{1.15}
\caption{Examples of Potential Risks for Stakeholders of MWBDs}
\label{tab:attackers_motivation}
\setlength{\tabcolsep}{3pt}
\small
\centering
\begin{tabular}{| m{50pt} | m{165pt} |}
\rowcolor[HTML]{E1E8F5}
\hline
\multicolumn{1}{|c|}{\textbf{Stakeholder}} 
    & \multicolumn{1}{c|}{\textbf{Potential Risk}}\\
\hline
\centering User &  \begin{enumerate*}[label=(\arabic*)]
    \item The corruption of medical data may lead to wrong diagnoses, and therefore wrong therapies. 
    \item Privacy leakage may result in stolen identities and electronic fraud.
    \item Denial of service (DoS) may harm the patient or even cause their death.
\end{enumerate*}\\
\hline
\centering Manufacturer & \begin{enumerate*}[label=(\arabic*)]
    \item The leakage of intellectual property (IP) may be used by competitors to increase their market share. 
    \item The leakage of IP may also lead to the increase in the number of counterfeits and attacks. 
    \item If an attack harms the patient, it may damage the manufacturer's reputation.
    \item Additionally, lawsuits may be filed against the manufacturer.\end{enumerate*}\\
\hline
\centering Hospital & If hospitals deal with corrupted MWBDs or counterfeits, they could (1)~lose their professional reputation and trust, (2)~be involved in lawsuits, or (3)~lose their accreditation.\\
\hline
\end{tabular}
\end{table}


\begin{figure}[t!]
    \centering
    \includegraphics[width=\columnwidth]{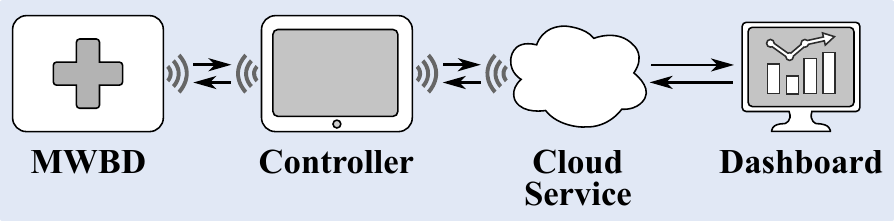}
    \captionsetup{justification=raggedright, singlelinecheck = false}
    \caption{A typical biomedical system. Adapted from~\cite{selimis2020security}.}
    \label{fig:biomedsystem}
    \vspace{-5mm}
\end{figure}

To formulate security objectives efficiently, it is necessary to have a perspective on the whole system. A typical next-generation biomedical system, outlined in Fig.~\ref{fig:biomedsystem}, consists of an MWBD wirelessly connected to an external controller. An MWBD usually has a limited functionality:
\begin{enumerate*} [label=(\arabic*)]
    \item it may serve as a controllable actuator, capable of processing a small set of external instructions from the controller; or
    \item it can operate as a smart sensor, which transmits the collected medical data to the controller.
\end{enumerate*}
The controller typically acts as a gateway and transfers the user health-related data to a cloud service for post-processing. The data processed in the cloud is then sent to a dashboard for the authorized users~\cite{selimis2020security}. The current biomedical systems implement protection mechanisms mainly beginning from the gateway~\cite{tuli2019next}. The lack of security in the MWBD puts the whole system in danger. Therefore, security should be built into MWBDs to protect their users~\cite{williams2016always}.

The MWBDs in the scope of this work typically have at least one network interface (wireless telemetry) and at least one transducer (sensor or actuator). These capabilities allow categorizing MWBDs as the Internet of Things (IoT) devices~\cite{wg_interagency_2018}. The National Institute of Standards and Technology (NIST) maintains a cybersecurity program dedicated to IoT~\cite{thelmaallennistgov_nist_2016}, reflecting the high demand for security for these devices. Some NIST standards applicable to MWBDs include NISTIR8200~\cite{wg_interagency_2018}, NISTIR8228~\cite{boeckl_considerations_2019}, NISTIR8259~\cite{fagan_foundational_2020}, and NISTIR8259A~\cite{fagan_iot_2020}. The developers of MWBDs should also be aware of the drafts from NIST~\cite{fagan_iot_2020_800-213, fagan_iot_2020_non_tech, fagan_creating_2020, fagan_profile_2020}, which will eventually be turned into standards.

While the manufacturers and designers of MWBDs may often not have direct experience with cybersecurity-related technologies~\cite{fagan_creating_2020}, implementating these mechanisms is of high importance and must be systematic. Accordingly, a standardized approach to profile potential attackers and to catalog potential threats is needed for the diverse architectures of emerging MWBDs, as the US Food and Drug Administration (FDA) announced recently~\cite{april_9_fdas_nodate}. The process of understanding, documenting, and evaluating system vulnerabilities, followed by addressing the protective measures, is known as threat modeling~\cite{torr2005demystifying}. By using threat modeling, it is possible to mitigate the weaknesses of systems against specific adversary scenarios in the early stages of the devices' life cycles~\cite{howard2002securecode}.

In this work, a methodology for modeling and semi-quantitatively assessing potential threats for the next-generation MWBDs is proposed. This work is primarily intended to inform the architects and manufacturers of new devices and pave the way for embedding security in MWBDs. The remainder of this publication is organized as follows. Section~\ref{sec:sec_challenges} provides the background on the security challenges in MWBDs. Section~\ref{sec:general_threat_modeling} presents some general considerations for the threat modeling process. Section~\ref{sec:threat_model} describes the proposed threat model for MWBDs. Section~\ref{sec:case_studies} presents an application of this model to real case studies of the four primary categories of MWBDs. Section~\ref{sec:discussion} discusses the results and makes suggestions for future work. Section~\ref{sec:conclusion} concludes the work. 
\section{Security Challenges for Emerging Miniaturized Wireless Biomedical Devices}
\label{sec:sec_challenges}

\noindent This chapter discusses various security challenges for emerging MWBDs. Some of these challenges are already present and require immediate attention. Other challenges are only anticipated in the future. However, both these categories of challenges illustrate the importance of security for MWBDs.
\vspace{-3mm}

\subsection{Limited Resources}
\noindent MWBDs are extremely constrained in area, weight, power, storage, network interfaces, and computing resources~\cite{williams2016always}. Therefore, they stand apart from classical information technology (IT) devices (e.g., smartphones, servers, or laptops), which have been used to define device cybersecurity capabilities~\cite{fagan_foundational_2020}. The limited resources available to MWBDs bound the range of security mechanisms applicable to these devices. Such as, the limited area excludes embedding complex security units occupying a lot of silicon on the chip. The constrained power excludes complex cryptographic computations and narrows the communication bandwidth and range. The restricted memory and performance prevent using sophisticated cryptographic algorithms~\cite{camara2015security}. Therefore, even though multiple modern cryptographic algorithms are reliable, the simplicity of MWBDs makes them unavailable for these devices~\cite{fagan_foundational_2020}. Correspondingly, lightweight cryptographic algorithms, suitable for constrained environments, need to be developed and standardized~\cite{alioto2019hwsec, camara2015security, computer_security_division_lightweight_2017}.

\vspace{-3mm}
\subsection{Multiple Attack Surfaces}
\noindent While being restricted in resources, MWBDs are often equipped with diverse transducers (sensors and actuators), and employ various communication (network) and power delivery techniques. All these interfaces may be considered potential channels via which an intruder might maliciously interact with the device~\cite{di2007hardware}. For a typical MWBD, five such channels may be identified. Three input channels include the control channel, the sensing channel, and the power delivering channel. Two output channels include the user data transferring channel and the actuating channel. Apart from the attacks on these five external interfaces, additional internal attack surfaces include the the memory attacks and the digital hardware. An attacker model covering mentioned channels was presented in~\cite{vakhter2020minimum}. 

\vspace{-3mm}
\subsection{Patient's Safety}
\noindent Being limited in resources and having multiple attack surfaces, MWBDs should use protective schemes that would not endanger a patient's life in an emergency~\cite{camara2015security}. Therefore, while these devices require server-side authentication to ensure that commands are authorized, the critical care services must be able to access the device even when the normal authentication method is unavailable. Hence, including the patient into the authentication schemes, like the one proposed in~\cite{9075945}, is potentially dangerous. Also, a direct disregarding of authentication and authorization in an emergency might introduce many potential threats. Because of that, authentication in medical devices remains an open problem~\cite{camara2015security}.

In general, for MWBDs, security and privacy requirements for a device should not affect its safety, reliability, and resiliency~\cite{boeckl_considerations_2019}. Traditional IT security prioritizes confidentiality, integrity, and availability. The ability of MWBDs to interact with the physical world through sensors and actuators requires addressing  threats to patients and their environments. Depending on the functionality of a particular biomedical device and its vital necessity for the patient, availability or integrity may be the highest priority, followed by privacy and finally confidentiality~\cite{wg_interagency_2018}.

\vspace{-3mm}
\subsection{Distributed Supply Chain}
\noindent The manufacturing of MWBDs relies on a complex, distributed supply chain. This chain contains multiple entities, distribution routes, technologies, as well as diverse legislation and practices. This multifariousness affects the design, fabrication, distribution, deployment, usage, and maintenance of MWBDs. Therefore, whether intentionally or unintentionally, the final users of MWBDs are at risk of supply chain attacks. Supply chain risks for MWBDs may include the insertion of malicious logic blocks, the use of unauthorized components and counterfeits, tampering, poor manufacturing and design practices, etc~\cite{oreilly_2019_2020}. Component suppliers often have poor cyber hygiene, and these vulnerabilities are more of an issue than the ingenuity of the attackers~\cite{megas_internet_2017}.

\vspace{-3mm}
\subsection{Lack of Incentives}
\noindent There is a lack of incentives to build security and privacy into IoT devices. Cybersecurity is often an afterthought to getting to market, with price and features prioritized. There is also a general lack of consumer education, leading to a lack of demand for better cybersecurity and privacy. There are guidelines available to help manufacturers mitigate risks, but a lack of incentives to adhere to them~\cite{megas_internet_2017, cardenas_cyber-physical_2019}.

\vspace{-3mm}
\subsection{Oncoming Challenges}
\noindent IT innovation is outpacing the development of supporting standards. With the changing threat environment, the cybersecurity needs of the future should be considered~\cite{wg_interagency_2018}. One such challenge for cryptography as a whole is that if large-scale quantum computers are ever built, many current public-key cryptosystems will be broken~\cite{oreilly_2019_2020}. That would compromise the information security of digital communication. Therefore, NIST initiated a process of post-quantum cryptography standardization, including quantum-resistant lightweight algorithms for resource-restricted devices  ~\cite{computer_security_division_post-quantum_2017, robinmateresenistgov_nist_2019}.

\vspace{4mm}
\noindent The list of challenges for MWBDs is not limited by the above examples. Any influx of new technologies will introduce new security challenges~\cite{franklin_security_2020}, and new countermeasures should be proposed accordingly. Security is never free and may cause an extra overhead (e.g., exceeding a tight power budget, increasing time delays, or causing extra memory usage, etc.)~\cite{camara2015security}. Considering all the limitations, it might be adequate to talk about the compromises between security and other parameters, when these biomedical devices contain, at least, some basic protective mechanisms against the most possible attacks. Lightweight security does not mean weak security. However, the lightweight security properties may be different from those desired for general use: it may be less robust, less misuse resistant, and have fewer features~\cite{wpi2020nist}.

The development of protective mechanisms and their embedding into devices require additional time and financial costs. Yet, ad-hoc security is insufficient for next-generation biomedical devices~\cite{vakhter2020minimum}. However, if developers follow well-articulated and transparent principles and practices, adding secure mechanisms into devices is repeatable~\cite{hunt2017seven}. Therefore, developers should have a guidance for the threat modeling process, allowing them to estimate and mitigate threats in the early stages of the device life cycle. In the next section, some general considerations about threat modeling process will be presented.              
\section{Threat Modeling Methodology}
\label{sec:general_threat_modeling}

\begin{figure}[h!]
    \centering
     \includegraphics[width=\columnwidth]{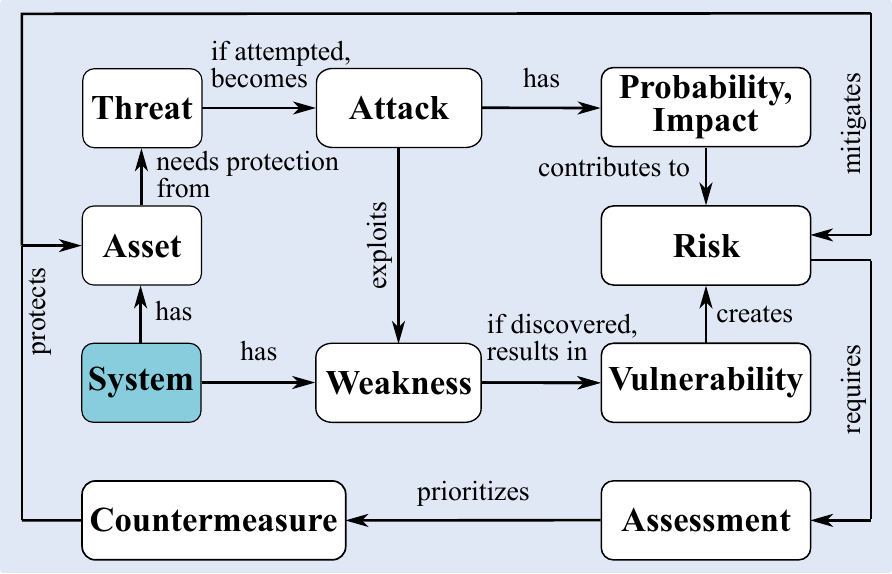}
    \captionsetup{justification=raggedright, singlelinecheck = false}
    \caption{Interconnection of terms in threat modeling and risk assessment. Adapted from~\cite{sanz2010threat}.}
    \label{fig:terms_interconnection}
\end{figure}

\noindent Considering the potential effects of cyber-attacks against the emerging MWBDs, it is necessary to plan for these intrusions and to take steps to prevent them~\cite{edge2020covid}. Therefore, a high-level method aimed to reveal, document, and address the security flaws of a system is demanded for these devices. This method is called threat modeling. Threat modeling uses special security terms, the main of which are assets, vulnerabilities, threats, attacks, risk, and risk assessment. Interconnection of these terms is shown in Fig.~\ref{fig:terms_interconnection}.

An asset is any item of value present in the system that must be kept secure and that an adversary aims to steal, modify, or disrupt~\cite{torr2005demystifying, vakhter2020minimum}. A vulnerability is a weakness in a system caused by a bad design or implementation. A threat is a circumstance or an event with the potential to have a malicious effect on assets, individuals, or organizations~\cite{initiative_guide_2012}. An attack is a malicious activity that attempts to threaten an asset by exploiting a vulnerability~\cite{howard2002securecode}. Risk is a measure of the extent to which an entity is threatened by a potential circumstance or an event, and typically is a function of \begin{enumerate*}[label=(\arabic*)]
    \item the adverse impacts that would arise if the circumstance or event occurs; and
    \item the likelihood of occurrence~\cite{initiative_guide_2012}.
    \end{enumerate*}
Risk assessment is the process of identifying, estimating, and prioritizing risks to assets, individuals, or organizations~\cite{initiative_guide_2012}. In particular, qualitative-quantitative risk assessment is a set of methods, principles, or rules for assessing risk based on the use of both qualitative terms and associated with them numbers~\cite{initiative_guide_2012}.

\begin{figure}[t!]
    \centering
    \includegraphics[width=\columnwidth]{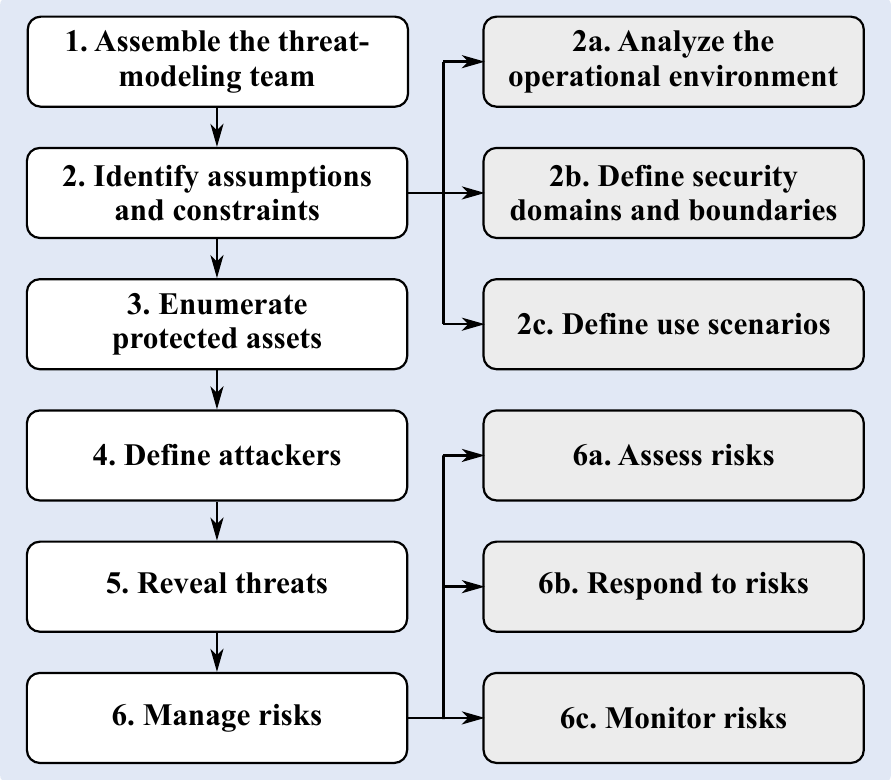}
    \captionsetup{justification=raggedright, singlelinecheck = false}
    \caption{Six primary steps of the threat modeling process. Adapted from~\cite{torr2005demystifying, initiative_guide_2012, howard2002securecode}.}
    \label{fig:threat_modeling_steps}
    \vspace{-5mm}
\end{figure}

Threat modeling is a multistage iterative process that provides insights on the assets that adversaries may be attracted by and allows recognizing the most probable attack vectors~\cite{cagnazzo2018threat, seeam2019threat, howard2002securecode}. The ultimate goal of threat modeling is to reduce the overall threat risk to an acceptable level. During threat modeling, all its steps should be collected and organized into a threat-model document~\cite{torr2005demystifying, arm2018threats, seeam2019threat}. This document should be kept current, reflecting new threats and mitigations as they originate~\cite{torr2005demystifying}. Threat modeling should be included in the overall development-and-documentation life-cycle~\cite{seeam2019threat}. Setting it apart from the overall design life-cycle may decrease the number of developers recognizing its importance~\cite{torr2005demystifying}.

The following subsections provide a high-level description of the primary phases of the threat modeling process, shown in Fig.~\ref{fig:threat_modeling_steps}. Activities highlighted for each phase build on the outcomes of prior activities. The steps provided for each phase are meant as a starting point and do not entirely define each activity.

\vspace{-3mm}
\subsection{Assemble the Threat-Modeling Team}
\noindent A team that will perform threat modeling should be assembled first. A threat modeling team should consist of at least one member from each engineering group (hardware, wireless link, software, and others) to guarantee a complete understanding of underlying technologies~\cite{howard2002securecode}.

\vspace{-3mm}
\subsection{Identify Assumptions and Constraints}
\noindent The next step in the process is to identify security assumptions and constraints under which the threat modeling is performed. It allows to capture the information in an appropriate level of abstraction. These assumptions must be verified later~\cite{torr2005demystifying}. This step includes three substeps:

\subsubsection{Analyze the Operational Environment}
\noindent the team captures the information about the infrastructure and describes the environment in which risk-based decisions are made. It helps to understand how different objects and elements in the system (an MWBD, a controller, a user, medical personnel, and other participants) interact with each other.

\subsubsection{Define Security Domains and Boundaries}
\noindent in this step, the primary logical components (for example, the analog front-end, the power management module, the data link, etc.) in the system are identified. Each logical component may be composed of several physical components, and have different entry points and threats~\cite{torr2005demystifying, di2007hardware}. Later, these logical components may be decomposed or merged to achieve a manageable level of granularity. For example, it may be appropriate to talk about the analog front-end of a device as a whole, or it may be essential to analyze the individual functional blocks that form this front-end.

No components are completely trusted, but rather various trust levels may be assigned to them (for example, high and low trusted components). After the high and low trusted logical components in the system are identified, the boundaries and interfaces between them should be determined~\cite{howard2002securecode}. After assumptions about the trust boundaries are made, threat analysis is usually performed for the data crossing these boundaries. The analysis must consider the direction of the data moving between trusted and untrusted components.

\subsubsection{Define Use Scenarios}
\noindent security measures are application-dependent~\cite{atamli2014threat}. For each system component, use scenarios provide a high-level description of how it will be implemented, deployed, and used. In order to better understand the system behavior, it may also be useful to list "anti-scenarios," which are the settings or the usage scenarios that are known to be vulnerable or restricted~\cite{torr2005demystifying, atamli2014threat}.

\vspace{-3mm}
\subsection{Enumerate Protected Assets}
\noindent At this stage, assets that will be protected should be identified and listed for the investigated design. Assets may be tangible (such as, user's personal information, therapies, or encryption keys) and intangible (for example, data consistency, data secrecy, data integrity, or data availability)~\cite{hasan_toward_2005}. The list of protected assets is later used in the risk analysis. This list for a particular design must be made considering the perspective of different stakeholders~\cite{vakhter2020minimum}. 

\vspace{-3mm}
\subsection{Define Attackers}
\noindent Understanding the attacker type is important to understand the resources and capabilities that they have at their disposal~\cite{hasan_toward_2005}. While real attackers rarely fit into one category, at a high level, they can be classified based on: \begin{enumerate*}[label=(\arabic*)] \item their position relative to the system (external or internal adversaries), \item their ability to intervene into the system (passive or active adversaries), \item their number (a single entity or a coordinated group), and \item the level of their expertise and equipment (sophisticated or unsophisticated)~\cite{rushanan2014sok, atamli2014threat, camara2015security}.\end{enumerate*}

An external intruder is an outside entity that is not a part of the system and does not have an authorized access~\cite{atamli2014threat}. An internal intruder may be: \begin{enumerate*}[label=(\alph*)] \item a malicious user who performs attacks to learn the secrets of the manufacturer or to get access to restricted functionality, and \item a bad manufacturer who has the ability to exploit the technology to collect information about the user or other devices~\cite{atamli2014threat}\end{enumerate*}. A passive eavesdropper is capable only to listen to the communication channel and to get access to the exchanged messages. These attackers are able to compromise the patient's privacy. They can determine if a person has a biomedical device; discover the type of device, its model and serial number; capture the information about the patient, such as the ID of their health records, name, age, diagnosis, therapy, and so forth~\cite{camara2015security}. An active adversary is not only capable to listen to the channel, but also to send or replay commands to the device and to modify or block messages. The motivation for the active attacks may be, for example, to cause malfunctioning or DoS to the device~\cite{camara2015security}.

\vspace{-3mm}
\subsection{Reveal Threats}
\noindent In the next step, threats will be revealed using systematic analysis. For example, threats can be identified by defining participants (like the user, the attacker, etc.), their actions, and the consequences of those actions~\cite{torr2005demystifying}. The objective is to enumerate the ways by which an attacker can compromise the system~\cite{howard2002securecode}. 

Threat modeling appears to be more productive when people have an understanding of how to attack systems~\cite{howard2002securecode}. For example, a kill-chain model~\cite{hutchins2011intelligence} studies intrusions from the adversaries' perspectives by incorporating the analysis of adversaries, their capabilities, objectives, attitudes, and limitations. In a kill-chain model, intrusions are described not as singular events but as phased progressions. This model illustrates that, in fact, the adversary must successfully move through each stage of the chain to achieve the desired goal. Therefore, just one mitigation breaks the chain and stops the adversary~\cite{hutchins2011intelligence}.

To help people remember the types of threats to which system components might be exposed~\cite{torr2005demystifying}, the STRIDE model~\cite{howard2002securecode} appears to be one of the most known models~\cite{shevchenko2018threat}. STRIDE is an abbreviation for spoofing, tampering, repudiation, information disclosure, denial of service, and elevation of privilege~\cite{howard2002securecode}. While STRIDE was developed for software systems, and MWBDs are predominantly hardware systems, some works~\cite{cagnazzo2018threat, seeam2019threat} adopted STRIDE to classify threats for mobile health systems and IoT devices. Yet, both of these works primarily focused on the security and privacy of data flows across networks rather than the specificity of MWBDs.

\vspace{-3mm}
\subsection{Manage Risks}
\noindent After the risks for a system are framed, risk management process should include:

\subsubsection{Assess Risks}
risk assessment is a crucial part of effective risk management. It is used to identify, estimate, and prioritize risks. The purpose of risk assessment is to inform decision-makers and to plan for risk responses. The result of risk assessment is a ranked list of threats that reflects the impact of attacks and the likelihood that harm will occur~\cite{initiative_guide_2012}.

Risk and its contributing factors can be assessed in a variety of ways, including quantitatively, qualitatively, or semi-quantitatively~\cite{initiative_guide_2012}. While both quantitative and qualitative assessments have their limitations, the semi-quantitative assessment provides the benefits of both these approaches. This method typically employs bins, scales, or representative numbers. Bins or scales translate easily into qualitative terms and also allow relative comparisons between values. The role of expert judgment in assigning values is more evident than in a purely quantitative approach. Also, when scales or sets of bins provide enough granularity, relative prioritization among results is better supported than in a purely qualitative approach~\cite{initiative_guide_2012}.

The DREAD model~\cite{howard2002securecode} was proposed to assess the associated risks for specific threats in mobile health systems~\cite{cagnazzo2018threat}. DREAD is an acronym for damage potential, reproducibility, exploitability, affected users, and discoverability. However, this model was initially designed to rank errors, flaws, or faults in software~\cite{howard2002securecode}. Since MWBDs are predominantly hardware systems, this model is not entirely suitable for them. For example, in the case of MWBDs, there is only one user, and therefore the criteria of affected users would not contribute to the final decision. The category of discoverability also seems not to be informative as it is difficult to estimate and is usually set equal to the maximum possible value~\cite{howard2002securecode}. 

The smartcard community introduced the guidance metrics to calculate the total effort required by an attacker to perform a successful attack. This guidance is described in the CCDB-2009-03-001, Common Criteria "Application of Attack Potential to Smartcards"~\cite{ccdb20090301}. This document aims to be applied to the operational behavior of a smartcard and applications specific only to hardware or software. Therefore, since MWBDs are mostly hardware systems, while some of them also have processing units running software, this document could serve as a base for the methodology to efficiently evaluate the threats for MWBDs.

\subsubsection{Respond to Risks}
in this step, the corresponding techniques and technologies should be chosen to respond to the discovered threats. Depending on the threat model, customers, and expected use cases, various countermeasures may be proposed~\cite{cybok-hs, fagan_iot_2020}. As a starting point for MWBDs, the model for the lightweight implementation of data security may be applied~\cite{vakhter2020minimum}. While providing a guideline on how to start protecting MWBDs, this model primarily focuses on data security. Therefore, other interfaces (sensing channel, power delivering channel, and actuating channel) require a separate analysis of available protective mechanisms. In general, engineers should weigh the value of each security countermeasure for MWBDs to reach a trade-off between safety, reliability, resilience, security, and privacy risks.

\subsubsection{Monitor Risks}
risk assessment is not simply a one-time activity that provides permanent and definitive information for decision-makers. Monitoring risk factors (threats, vulnerabilities, capabilities and intent of adversaries, etc.) over time can provide critical information on changing conditions that could potentially affect the security of systems. Information derived from the ongoing risk monitoring can be used to refresh risk assessments~\cite{initiative_guide_2012}.

In the next chapter, a specific threat model for emerging MWBDs will be introduced. According to the described threat modeling methodology, assumptions about the operational environment, security boundaries, and use scenarios will be identified first. Then, suggestions about protected assets and attackers will be provided. Finally, a risk assessment methodology will be proposed.              
\section{Proposed Threat Model for MWBDs}
\label{sec:threat_model}

\begin{figure}[t!]
    \centering
    \includegraphics[width=\columnwidth]{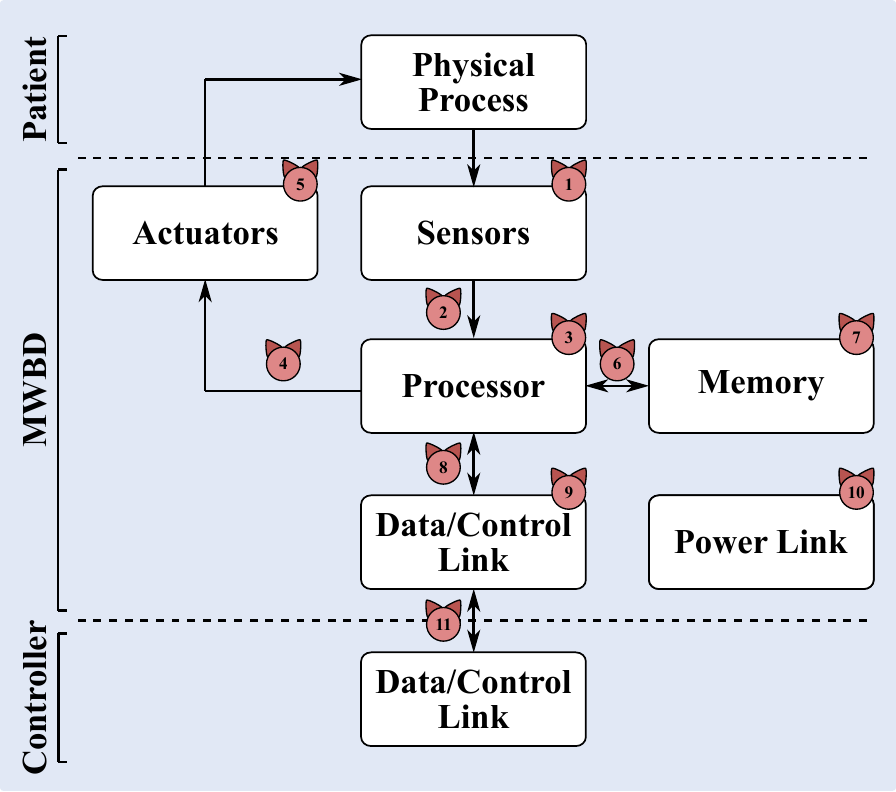}
    \captionsetup{justification=raggedright, singlelinecheck = false}
    \caption{General architecture of an MWBD. Attack points are shown in red. Adapted from~\cite{cardenas_cyber-physical_2019}.}
    \label{fig:attack_points}
\end{figure}

\noindent While at least a dozen of threat modeling methods exist in IT~\cite{shevchenko2018threat}, they have mostly been developed for either purely software systems or networking systems. However, MWBDs are predominantly hardware systems, and these existing threat modeling techniques may be less efficient for them. Therefore, describing a specific threat model for emerging MWBDs would help developers to make their systems more secure.

The generic models for implementing security in MWBDs, like the one proposed in~\cite{vakhter2020minimum}, may be a good starting point. However, each particular device would require performing a separate threat analysis. The design of a threat model requires security expertise. After the threat model is defined, the threat analysis becomes an engineering task and may be performed by non-experts in security~\cite{howard2002securecode}. 

We propose a domain-specific threat model that aims to help the designers and manufacturers of MWBDs to identify threats and embed security in their designs in the pre-market phase of the lifecycle of an MWBD. While some of the proposed threat modeling process components may or may not be applicable for a particular MWBD, the overall process is valid for a wide range of devices.

\vspace{-3mm}
\subsection{Assumptions}
\subsubsection{Operational Environment}
this model considers a single victim using an MWBD in a public space, accessible to multiple people, including but not limited to adversaries. Being in a public space, attackers can neither have physical access to the user (device) nor utilize large high-end equipment. For each particular case study of MWBDs, additional assumptions about its operational environment may be required.

\subsubsection{Security Domains and Boundaries}
in general, an MWBD controls and monitors some physical process (a health condition) in the human body. A set of sensors report the state of this health condition to the processor. Based on the information received from sensors, the processor defines the control signals to actuators to maintain the desired state. The processor often communicates with an external controller that monitors or configures the device. This communication is performed via a wireless telemetry interface (data/control link). The power is typically also delivered wirelessly, except in more complex battery-powered wearables or implantables, like pacemakers. The processor may store various sensitive data (end users' private data, the chip ID, etc.) in the off-chip memory. This general architecture, presented in Fig.~\ref{fig:attack_points}, considers eleven attack points, for which examples are provided in Table~\ref{tab:attack_points}.

\begin{table}[h!]
\renewcommand{\arraystretch}{1.15}
\caption{Attack Points - Examples of Attacks}
\label{tab:attack_points}
\centering
\setlength{\tabcolsep}{3pt}
\small
\begin{tabular}{| m{50pt} | m{165pt} |}
\rowcolor[HTML]{E1E8F5}
\hline
\multicolumn{1}{|c|}{\textbf{Attack point}} 
    & \multicolumn{1}{c|}{\textbf{Attack example}}\\ \hline
\centering 1 & Fault injection attack~\cite{6655254}. \\ \hline
\centering 2, 4, 6, 8 & Probing~\cite{aga_invisimem_2017}. \\ \hline
\centering 3 & Hardware trojans~\cite{7829523}. \\ \hline
\centering 5 & Control spoofing~\cite{hasan2019protecting}. \\ \hline
\centering 7 & Microprobing~\cite{skorobogatov_how_2017}. \\ \hline
\centering 9 & Shielding / cutting the antenna. \\ \hline
\centering 10 & \begin{tabular}[c]{@{}l@{}}Denial of sleep~\cite{brownfield_wireless_2005}. Power Analysis~\cite{zhang_power_2021}.\end{tabular} \\ \hline
\centering 11 & Man-in-the-middle (MIM)~\cite{lahmadi_mitm_2020}. \\ \hline
\end{tabular}
\end{table}

This work focuses on non-invasive direct attacks because the direct channels, missing security mechanisms, should be protected in the first place. Among the eleven defined attack points, direct interfaces include sensors (attack point 1), actuators (attack point 5), and telemetry (attack points 9, 10, and 11). For MWBDs, sensors and actuators are mainly located on/in the body. Therefore, it is hard to access them once they are deployed. Also, before deployment, there is a rigorous calibration for MWBDs, which makes it harder to deploy a tampered device. Therefore, for direct channels, attacks against telemetry (wireless data and power transfer) are prioritized in this work. Attack points 2 - 4 and 6 - 8 are out of scope for this model. Securing the remaining elements of the biomedical system (the physical process, the controller, the cloud, and the dashboard) needs separate analysis and is out of scope for this study.

To have a manageable level of complexity in the model, we consider that integrated circuits (ICs) are trustable after fabrication and testing, which means that malicious logic blocks are not inserted into the design (a chip manufacturer is trustable). After deployment, the IC design is fixed, in other words, there are no dynamic attacks on hardware. Semi-invasive (chip imaging, laser probing, voltage contrast, photo-emission, etc.) and invasive (reverse engineering, laser fault injection, etc.) hardware attacks are out of scope for this work.

\renewcommand{\arraystretch}{1.15}
\begin{table*}[t!]
\centering
\caption{Risk Assessment - Characteristics and Scales}
\label{tab:risk_assessment}
\resizebox{\textwidth}{!}{
\small
\begin{tabular}{| m{4pt} | l | m{38pt} |c|l|}
\hline
\rowcolor[HTML]{E1E8F5}
\multicolumn{1}{|c|}{\cellcolor[HTML]{FFFFFF}} & \multicolumn{1}{c|}{\cellcolor[HTML]{E1E8F5}Characteristic} & \multicolumn{1}{|c|}{QLV} & \begin{tabular}[c]{@{}c@{}}QNV\end{tabular} & \multicolumn{1}{c|}{\cellcolor[HTML]{E1E8F5}Description}
\\ \hline

\cellcolor[HTML]{E1E8F5} &  & Expert & 1 & \begin{tabular}[c]{@{}l@{}}Broad expertise in cybersecurity. Familiar with the target device at the developer level. Experienced \\ with, and equipped by, sophisticated tools, for which the expertise in using is difficult to obtain.\end{tabular} \\ \cline{3-5} 
\cellcolor[HTML]{E1E8F5} &  & Proficient & 2 & \begin{tabular}[c]{@{}l@{}}Familiar with security behavior, classical attacks~\cite{ccdb20090301}, and related disciplines (electrical engineering, \\ software development, etc.).\end{tabular} \\ \cline{3-5}

\cellcolor[HTML]{E1E8F5} & \multirow{-3}{*}[1em]{\begin{tabular}[c]{@{}l@{}}\textbf{C1: Expertise} \\ \textbf{of the attacker}\end{tabular}} & Layman & 3 & No particular expertise. \\ \cline{2-5} 
\cellcolor[HTML]{E1E8F5} &  & Custom & 1 & Bespoke equipment. \\ \cline{3-5} 
\cellcolor[HTML]{E1E8F5} &  & Specialized & 2 & \begin{tabular}[c]{@{}l@{}}Expensive commercially available equipment. Sales are controlled by manufacturers. The expertise \\ in using the equipment is difficult to obtain. For example, the type of expensive equipment which \\ universities have in their possession~\cite{ccdb20090301}.\end{tabular} \\ \cline{3-5}

\cellcolor[HTML]{E1E8F5} & \multirow{-3}{*}[1em]{\begin{tabular}[c]{@{}l@{}}\textbf{C2: Equipment} \\ \textbf{required to carry} \\ \textbf{out the attack}\end{tabular}} & Standard & 3 & \begin{tabular}[c]{@{}l@{}}Mass-market commercially available equipment. The expertise in using the equipment may be \\ acquired from publicly available resources. For example, smartphones or laptops.\end{tabular} \\ \cline{2-5} 
\cellcolor[HTML]{E1E8F5} &  & Nearby & 1 & \begin{tabular}[c]{@{}l@{}}The attacker is in close proximity (the same room) to, and in immediate visibility to, the victim. \\ No physical obstacles, like walls and doors, exist between the attacker and the victim (device).\end{tabular} \\ \cline{3-5}

\cellcolor[HTML]{E1E8F5} &  & Moderate & 2 & \begin{tabular}[c]{@{}l@{}}The attacker is in the same space with the victim, there are no physical obstacles between the attacker \\ and the victim, but the distance between the attacker and the victim does not allow the victim to see \\ the attacker.\end{tabular} \\ \cline{3-5} 

\cellcolor[HTML]{E1E8F5} & \multirow{-3}{*}[2.2em]{\begin{tabular}[c]{@{}l@{}}\textbf{C3: Physical} \\ \textbf{proximity to} \\ \textbf{the attacked} \\ \textbf{device}\end{tabular}} & Remote & 3 & The attacker is capable to mount an attack while in a different location than the victim. \\ \cline{2-5} 
\cellcolor[HTML]{E1E8F5} &  & Long & 1 & The attacker is able to access the device continuously. \\ \cline{3-5} 
\cellcolor[HTML]{E1E8F5} &  & Moderate & 2 & The attacker is able to access the device multiple times. \\ \cline{3-5} 

\cellcolor[HTML]{E1E8F5} & \multirow{-3}{*} {\begin{tabular}[c]{@{}l@{}}\textbf{C4: Device} \\ \textbf{access time}\end{tabular}} & Short & 3 & The attacker is able to access the device once in real time. \\ \cline{2-5} 
\cellcolor[HTML]{E1E8F5} &  & Critical & 1 & Low-level information about hardware design or source code~\cite{ccdb20090301}. \\ \cline{3-5} 
\cellcolor[HTML]{E1E8F5} &  & Restricted & 2 & Proprietary confidential developer's information like specifications or guidances~\cite{ccdb20090301}. \\ \cline{3-5} 

\multirow{-15}{*}[4em]{\cellcolor[HTML]{E1E8F5}\rotatebox[origin=c]{90}{Probability}} & \multirow{-3}{*}{\begin{tabular}[c]{@{}l@{}}\textbf{C5: Device} \\ \textbf{information} \end{tabular}} & Public & 3 & Public domain information. \\ \hline
\cellcolor[HTML]{E1E8F5} &  & High & 3 & \begin{tabular}[c]{@{}l@{}}Attacks have a severe or catastrophic adverse effect on the user. In some circumstances, DoS attacks \\ may lead to an irreparable harm, such as apoplexy, or even a loss of human life.\end{tabular} \\ \cline{3-5} 
\cellcolor[HTML]{E1E8F5} &  & Moderate & 2 & \begin{tabular}[c]{@{}l@{}}Attacks have a moderate adverse effect on the user. For example, alternating the sensor data may cause \\ wrong commands to actuators, directly impacting the health of the victim, but the malicious effect is \\ temporary.\end{tabular} \\ \cline{3-5} 

\multirow{-3}{*}[1.4em]{\cellcolor[HTML]{E1E8F5}\rotatebox[origin=c]{90}{Impact}} & \multirow{-3}{*}[1.7em]{\begin{tabular}[c]{@{}l@{}}\textbf{C6: Severity}\\ \textbf{of the attack}\end{tabular}} & Low & 1 & \begin{tabular}[c]{@{}l@{}}Attacks have a limited adverse effect on the user (pose surmountable problems for the victim). For \\ example, under certain context, loss of personal information do not prevent the MWBD from its correct \\ functioning.\end{tabular} \\ \hline

\multicolumn{5}{l}{Note: QLV and QNV stand for a qualitative value and a quantitative value accordingly.}
\end{tabular}
}
\end{table*}

\subsubsection{Use Scenarios}
use scenarios are unique for each device, and therefore cannot be generalized. Each particular MWBD would require listing its use scenarios. To give a clear picture to the readers through examples, Section~\ref{sec:case_studies} of this work provides case studies for each category of MWBDs, namely injectables, implantables, ingestibles, and wearables.

\vspace{-3mm}
\subsection{Protected Assets}
\noindent Among various stakeholders  of MWBDs, this model focuses on the user. This perspective requires balancing safety, service availability, resilience, and privacy. Safety protects from hazards, risks, or injury caused by the operation of the device. Service availability protects against denial of device service. Resilience means security against most attacks and the ability to return to a safe state in case of a successful attack. Privacy means protecting the confidentiality and integrity of personally identifiable information (PII).
Privacy goals include:
\begin{enumerate*}[label=(\arabic*)]
    \item device-existence privacy;
    \item device-type privacy;
    \item unique device ID privacy;
    \item measurement and log privacy;
    \item patient privacy; and
    \item patient location privacy~\cite{rushanan2014sok}. 
\end{enumerate*}
Confidentiality prevents the improper disclosure of information, and data integrity prevents the improper modification of information. 

Assets are unique for each MWBD. Therefore, each case study of MWBDs would require specific analysis. After the assets are listed, attacks for each of them should be defined.

\vspace{-3mm}
\subsection{Attackers}
\noindent This model considers sophisticated attackers, who have the intent and capabilities to attack the MWBD. The attacker may be both an individual (outsider, insider, trusted insider, or privileged insider) and an established group. The attacker may be both passive and active. The adversary, however, does not have physical access to the user of the MWBD, and hence all attacks are remote. For this reason, attacks on the physical process are not considered in the proposed model. However, for the smartwatch-like wearables, the attacker may manipulate the device for a limited period of time before the user begins to use it (if the device was left unattended after deployment). Large high-end equipment is excluded from consideration in this paper since an attacker is not able to bring it to a public place.

\renewcommand{\arraystretch}{1.15}
\begin{table*}[h!]
\caption{Case Studies - Devices and Assets}
\label{tab:case_studies_devices}
\resizebox{\textwidth}{!}{%
\small
\begin{tabular}{|c|c|c|l|l|l|c|}
\hline
\rowcolor[HTML]{E1E8F5} 
Number & Name & Category & \multicolumn{1}{c|}{\cellcolor[HTML]{E1E8F5}Purpose} & \multicolumn{1}{c|}{\cellcolor[HTML]{E1E8F5}Status} & \multicolumn{1}{c|}{\cellcolor[HTML]{E1E8F5}Assets} & Ref. \\ \hline

D1 & BioMote & Injectable & \begin{tabular}[c]{@{}l@{}}A wireless sensor node for \\ continuous monitoring of the\\ blood alcohol content \\ (ethanol, background, and pH).\end{tabular} & \begin{tabular}[c]{@{}l@{}}In-vitro\\ tests.\end{tabular} & \begin{tabular}[c]{@{}l@{}}1. Backscattered user sensor data\\ (blood alcohol content).\end{tabular} & \cite{jiang2019biomote} \\ \hline

D2 & \begin{tabular}[c]{@{}c@{}}Wireless\\ Capsule Endoscope\end{tabular} & Ingestible & \begin{tabular}[c]{@{}l@{}}A wireless spherical endoscopic \\ capsule for ColoRectal Cancer (CRC) \\ screening with a locomotion control.\end{tabular} & \begin{tabular}[c]{@{}l@{}}In-vitro\\ tests.\end{tabular} & \begin{tabular}[c]{@{}l@{}}1. User data (cancer information,\\ video frames)\\ 2. Control signals from an external\\ device.\end{tabular} & \cite{fontana2016innovative} \\ \hline

D3 & \begin{tabular}[c]{@{}c@{}}Trimodal wireless\\ implantable\\ neural interface\\ System-on-Chip (SoC)\end{tabular} & Implantable & \begin{tabular}[c]{@{}l@{}}A wireless trimodal neural interface \\ SoC, providing optical/electrical \\ stimulation capabilities and neural\\ recordings.\end{tabular} & \begin{tabular}[c]{@{}l@{}}In-vivo\\ tests\\ in freely\\ behaving\\ animals.\end{tabular} & \begin{tabular}[c]{@{}l@{}}1. Recording/stimulation \\ parameters (BLE and on-off-keying \\ (OOK) signals at 13.56~MHz).\\ 2. Evoked neural activities \\ (OOK RF signals at 433~MHz).\end{tabular} & \cite{jia2020soc} \\ \hline

D4 & \begin{tabular}[c]{@{}c@{}}An integrated readout \\ circuit for a transcutaneous \\ oxygen sensing wearable \\ device\end{tabular} & Wearable & \begin{tabular}[c]{@{}l@{}}A fluorescence-based readout \\ dedicated to sensing transcutaneous \\ oxygen diffusing through the skin.\end{tabular} & \begin{tabular}[c]{@{}l@{}}Ex-vivo\\ tests.\end{tabular} & \begin{tabular}[c]{@{}l@{}}1. User data (partial pressure \\ of transcutaneous $O_2$).\\ 2. Control signals from \\ an external controller.\end{tabular} & \cite{costanzo2020integrated} \\ \hline
\end{tabular}
}
\end{table*}

\vspace{-3mm}
\subsection{Risk Assessment}
\noindent In the proposed model, six relevant characteristics of threats are selected to assess risks for an MWBD (see Table~\ref{tab:risk_assessment}):
\begin{enumerate}[label=\arabic*.]
    \item \textit{Expertise of the attacker} reflects the extent to which related knowledge is necessary for the adversary to perform a successful attack.
    \item \textit{Equipment required to carry out the attack} describes the tools that an adversary needs to use to carry out an attack.
    \item \textit{Physical proximity to the attacked device} shows how close the adversary should be to the user of an MWBD to mount a successful attack.
    \item \textit{Device access time} evaluates the time during which the attacker can have access to the attacked device.
    \item \textit{Device information} evaluates the need for the particular information assisting the attack, which cannot be substituted by a related combination of time and expertise~\cite{ccdb20090301}.
    \item \textit{Severity of the attack} estimates the loss caused by its occurrence. In this model, severity corresponds to the physical harm for the user caused by a successful attack.
\end{enumerate}

For each characteristic $C_1-C_6$ in Table~\ref{tab:risk_assessment}, a three-tiered qualitative-quantitative scale is assigned. While this scale may provide limited granularity, it makes the first iteration of the proposed model less complex and easier to apply. Additional tiers may be added when required.

Characteristics $C_1-C_5$ in Table~\ref{tab:risk_assessment} define the probability of an attack. The quantitative values for these categories are assigned in the reverse order: the fewer efforts are needed to perform an attack, the higher its likelihood, and therefore the higher the corresponding score. 
The total probability $P$ of an attack equals the sum of the values for $C_1 - C_5$.
The impact of an attack $I$ equals the value of the characteristic $C_6$ in Table~\ref{tab:risk_assessment}, for which the highest score corresponds to the highest severity.

After potential threats are identified, and their principal characteristics are captured, the total risk $R$ for each of them may be defined based on their impact $I$ and probability $P$. In risk management, the risk matrix approach is a typical qualitative-quantitative tool to evaluate various risks. Even though it is not mathematically rigorous, its visibility and ease of
application make it well-received in various industries~\cite{ni_extensions_2010}. 

\begin{table}[h!]
\renewcommand{\arraystretch}{1.15}
\caption{Risk matrix}
\label{tab:risk_matrix}
\centering
\begin{tabular}{|
>{\columncolor[HTML]{E1E8F5}}c |c|c|c|}
\hline
\cellcolor[HTML]{E1E8F5} & \multicolumn{3}{c|}{\cellcolor[HTML]{E1E8F5}\textbf{Impact (\textit{I})}} \\ \cline{2-4} 
\multirow{-2}{*}{\cellcolor[HTML]{E1E8F5}\textbf{Probability (\textit{P})}} & \cellcolor[HTML]{E1E8F5}\textbf{Low (1)} & \cellcolor[HTML]{E1E8F5}\textbf{Moderate (2)} & \cellcolor[HTML]{E1E8F5}\textbf{High (3)} \\ \hline
\textbf{Low (5-7)} & \cellcolor[HTML]{00B0F0}Very Low & \cellcolor[HTML]{F8FF00}Moderate & \cellcolor[HTML]{F8A102}High \\ \hline
\textbf{Moderate (8-12)} & \cellcolor[HTML]{32CB00}Low & \cellcolor[HTML]{F8FF00}Moderate & \cellcolor[HTML]{F8A102}High \\ \hline
\textbf{High (13-15)} & \cellcolor[HTML]{F8FF00}Moderate & \cellcolor[HTML]{F8A102}High & \cellcolor[HTML]{FD6864}Very High \\ \hline
\end{tabular}
\end{table}

\vspace{-1mm}

For this model, the risk matrix shown in Table~\ref{tab:risk_matrix} is used to assess risks, where different colors code different levels of risk. After the risk assessment is completed for each threat and the data are filled in the risk matrix, the results appear sorted according to their risk levels. While these results suggest which threats are perilous and require more attention, designers should decide which threat to address first, based on their abilities and specific requirements for the design.              
\section{Case Studies for Emerging Miniaturized Wireless Biomedical Devices}
\label{sec:case_studies}

\renewcommand{\arraystretch}{1.15}
\begin{table*}[t!]
\caption{Case Studies - Threats}
\label{tab:case_studies_threats}
\resizebox{\textwidth}{!}{%
\begin{tabular}{|c|l|c|c|c|c|c|c|c|c|c|}
\hline
\rowcolor[HTML]{E1E8F5} 
\cellcolor[HTML]{E1E8F5} & \multicolumn{1}{c|}{\cellcolor[HTML]{E1E8F5}} & \cellcolor[HTML]{E1E8F5} & \multicolumn{1}{l|}{\cellcolor[HTML]{E1E8F5}} & \multicolumn{6}{c|}{\cellcolor[HTML]{E1E8F5}Probability} & Impact \\ \cline{5-11} 
\rowcolor[HTML]{E1E8F5} 
\multirow{-2}{*}{\cellcolor[HTML]{E1E8F5}Number} & \multicolumn{1}{c|}{\multirow{-2}{*}{\cellcolor[HTML]{E1E8F5}Threat}} & \multirow{-2}{*}{\cellcolor[HTML]{E1E8F5}Threat violates} & \multicolumn{1}{l|}{\multirow{-2}{*}{\cellcolor[HTML]{E1E8F5}Device}} & C1 & C2 & C3 & C4 & C5 & Total & C6 \\ \hline

T1 & \begin{tabular}[c]{@{}l@{}}The attacker uses a counterfeit wearable controller to power up the\\ device and to collect the sensitive private health data. As a result,\\ the confidentiality of the patient's personal information is violated.\end{tabular} & \begin{tabular}[c]{@{}c@{}}Confidentiality\\ Authenticity\end{tabular} & D1 & 3 & 3 & 2 & 3 & 3 & \begin{tabular}[c]{@{}c@{}}14\\ (High)\end{tabular} & \begin{tabular}[c]{@{}c@{}}1\\ (Low)\end{tabular} \\ \hline

T2 & \begin{tabular}[c]{@{}l@{}}The attacker conducts a MIM attack using special equipment to tamper \\ with the data / control signals, producing a false report about the health \\ condition / a false command, causing a false treatment or therapy. This \\ may result in temporary or permanent health damage. Even if the device \\ does not have actuators, a physician working with the corrupted sensor \ \\data can prescribe a wrong treatment or therapy.\end{tabular} & Integrity & \begin{tabular}[c]{@{}c@{}}D1\\ D3\end{tabular} & 1 & 2 & 2 & 3 & 3 & \begin{tabular}[c]{@{}c@{}}11\\ (Moderate)\end{tabular} & \begin{tabular}[c]{@{}c@{}}3\\ (High)\end{tabular} \\ \hline

T3 & \begin{tabular}[c]{@{}l@{}}The attacker jams the wireless data link. Sensor data cannot be \\ collected accurately, and stimulation cannot be applied correctly.\\ This may result in permanent health damage due to the incorrect\\ or missing treatment.\end{tabular} & Availability & \begin{tabular}[c]{@{}c@{}}D1\\ D3\end{tabular} & 1 & 2 & 3 & 2 & 3 & \begin{tabular}[c]{@{}c@{}}11\\ (Moderate)\end{tabular} & \begin{tabular}[c]{@{}c@{}}3\\ (High)\end{tabular} \\ \hline

T4 & \begin{tabular}[c]{@{}l@{}}The attacker uses a software-defined radio or an external hub to \\ collect the data about the evoked neural activities of the user. This \\ leads to a leak of the patient's personal confidential information.\end{tabular} & \begin{tabular}[c]{@{}c@{}}Confidentiality\\ Authenticity\end{tabular} & D3 & 2 & 3 & 2 & 3 & 3 & \begin{tabular}[c]{@{}c@{}}13\\ (High)\end{tabular} & \begin{tabular}[c]{@{}c@{}}1\\ (Low)\end{tabular} \\ \hline

T5 & \begin{tabular}[c]{@{}l@{}}The attacker eavesdrops the data, using the standard Bluetooth\\ equipment. This leads to a leak of the patient's confidential \\ information.\end{tabular} & \begin{tabular}[c]{@{}c@{}}Confidentiality\\ Authenticity\end{tabular} & \begin{tabular}[c]{@{}c@{}}D2\\ D4\end{tabular} & 2 & 2 & 2 & 3 & 3 & \begin{tabular}[c]{@{}c@{}}12\\ (Moderate)\end{tabular} & \begin{tabular}[c]{@{}c@{}}1\\ (Low)\end{tabular} \\ \hline

T6 & \begin{tabular}[c]{@{}l@{}}The attacker interferes with the communication channel and\\ substitutes the user data by some counterfeit data. This may result\\ in permanent health damage due to a wrong or missing therapy.\end{tabular} & Authenticity & \begin{tabular}[c]{@{}c@{}}D2\\ D4\end{tabular} & 1 & 2 & 2 & 3 & 3 & \begin{tabular}[c]{@{}c@{}}11\\ (Moderate)\end{tabular} & \begin{tabular}[c]{@{}c@{}}3\\ (High)\end{tabular} \\ \hline

T7 & \begin{tabular}[c]{@{}l@{}}The attacker replays a command to decrease the illumination or\\ to switch the Bluetooth module off to decrease the amount of\\ information that can be extracted from video frames. This would\\ require to repeat the measurements with a different device.\end{tabular} & Availability & D2 & 2 & 2 & 2 & 3 & 3 & \begin{tabular}[c]{@{}c@{}}12\\ (Moderate)\end{tabular} & \begin{tabular}[c]{@{}c@{}}1\\ (Low)\end{tabular} \\ \hline
 &  &  & D2 &  &  &  &  &  &  & \begin{tabular}[c]{@{}c@{}}1\\ (low)\end{tabular} \\ \cline{4-4} \cline{11-11} 

\multirow{-2}{*}[0.5em]{T8} & \multirow{-2}{*}[1.8em]{\begin{tabular}[c]{@{}l@{}}The attacker sends a high-volume radio traffic to deplete the battery.\\ Even if the device supports authentication, the process of commands'\\ and data validation would consume extra power, which could lead\\ eventually to denial of service.  \end{tabular}} & \multirow{-2}{*}[0.5em]{Availability} & D4 & \multirow{-2}{*}[0.5em]{2} & \multirow{-2}{*}[0.5em]{2} & \multirow{-2}{*}[0.5em]{2} & \multirow{-2}{*}[0.5em]{3} & \multirow{-2}{*}[0.5em]{3} & \multirow{-2}{*}{\begin{tabular}[c]{@{}c@{}}12\\ (Moderate)\end{tabular}} & \begin{tabular}[c]{@{}c@{}}3\\ (High)\end{tabular} \\ \hline

\multicolumn{11}{l}{
\begin{tabular}[c]{@{}l@{}} Note: for T8, D2 and D4 have different impacts. D2 is used for short-term monitoring of the digestive tract. Measurements are performed in a laboratory under the \\ supervision of a physician. In this case, the denial of service may be quickly resolved and the procedure may be repeated with another device. However, D4 can be used \\ outside the hospital. In the event of DoS, it may not be quickly replaced with another device. Therefore, the impact of T8 is higher for D4.\end{tabular}
}

\end{tabular}
}
\end{table*}

\noindent This chapter aims to provide meaningful examples of MWBDs (see Table~\ref{tab:case_studies_devices}) and their associated threats (see Table~\ref{tab:case_studies_threats}), which are disclosed and prioritized with the application of the proposed threat model. For each category of MWBDs, one representative device was selected. At the time of writing this paper, most MWBDs were either at the proof-of-concept stage or the stage of pre-clinical trials in freely-behaving animals. Nevertheless, since they have a potential for large-scale manufacturing and are intended to be ultimately used in humans, it is of interest to analyze threats for them using the proposed threat model.

The following subsections will provide brief overviews of these case studies, including information about their purposes, internal structure, and operational environments. For each of them, risk assessment results will be presented in the form of risk matrices. Since the primary purpose of this chapter is to illustrate the application of the designed model but not to perform a comprehensive threat modeling for the selected devices, it does not guarantee to include all possible threats for the selected case studies.

\vspace{-3mm}
\subsection{Case study 1 - Injectable}

\begin{figure}[h!]
    \centering
    \includegraphics[width=\columnwidth]{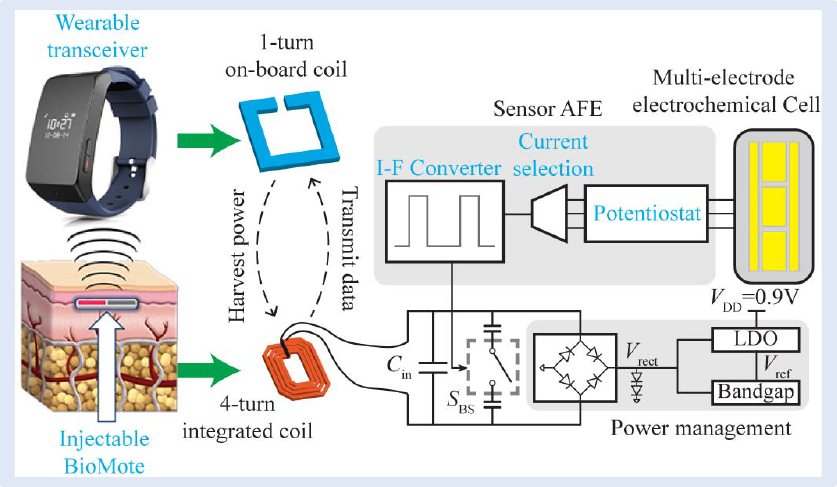}
    \captionsetup{justification=raggedright, singlelinecheck = false}
    \caption{Device 1: block diagram. Adapted from~\cite{jiang2019biomote}.}
    \label{fig:device_1}
\end{figure}

\noindent BioMote~\cite{jiang2019biomote}, shown in Fig.~\ref{fig:device_1} and further referred to as device~1 (D1), is a wireless sensor node intended for continuous monitoring of the blood alcohol content. D1 is routinely used outside of a clinical laboratory. D1 is subcutaneously injected into the interstitial fluid and wirelessly paired with an external wearable controller. The D1's electrochemical sensor array measures alcohol and pH. These measurements are unidirectionally transmitted to the controller through backscatter using a current-to-frequency converter. D1 is wirelessly powered by the controller via an inductive link.

\begin{table}[h!]
\renewcommand{\arraystretch}{1.15}
\centering
\caption{Risk matrix - Device 1}
\label{tab:risk_matrix_d1}
\begin{tabular}{|
>{\columncolor[HTML]{E1E8F5}}c |c|c|c|}
\hline
\cellcolor[HTML]{E1E8F5} & \multicolumn{3}{c|}{\cellcolor[HTML]{E1E8F5}\textbf{Impact (\textit{I})}} \\ \cline{2-4} 
\multirow{-2}{*}{\cellcolor[HTML]{E1E8F5}\textbf{Probability (\textit{P})}} & \cellcolor[HTML]{E1E8F5}\textbf{Low} & \cellcolor[HTML]{E1E8F5}\textbf{Moderate} & \cellcolor[HTML]{E1E8F5}\textbf{High} \\ \hline
\textbf{Low} & \cellcolor[HTML]{00B0F0} -- & \cellcolor[HTML]{F8FF00} -- & -- \cellcolor[HTML]{F8A102} \\ \hline
\textbf{Moderate} & \cellcolor[HTML]{32CB00} -- & \cellcolor[HTML]{F8FF00} -- & \cellcolor[HTML]{F8A102}\begin{tabular}[c]{@{}c@{}} T2, T3 \end{tabular} \\ \hline
\textbf{High} & \cellcolor[HTML]{F8FF00} T1 & \cellcolor[HTML]{F8A102}\begin{tabular}[c]{@{}c@{}} -- \end{tabular} & \cellcolor[HTML]{FD6864} -- \\ \hline
\end{tabular}
\end{table}

For D1, the risk matrix is presented in Table~\ref{tab:risk_matrix_d1}. It shows that T1 has a moderate risk while T2 and T3 have high risks.

\subsection{Case study 2 - Ingestible}

\begin{figure}[h!]
    \centering
    \includegraphics[width=\columnwidth]{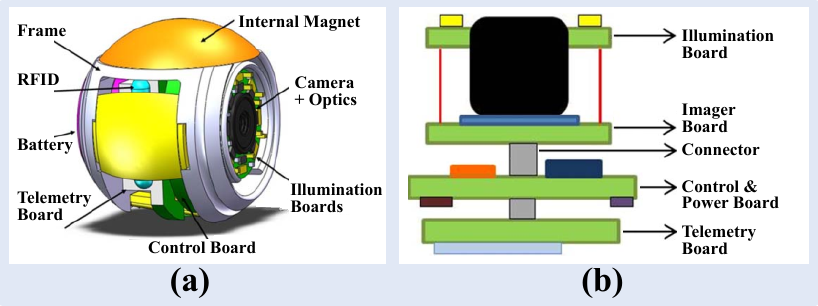}
    \captionsetup{justification=raggedright, singlelinecheck = false}
    \caption{Device 2: (a) 3D design. (b) Block diagram. Adapted from~\cite{fontana2016innovative}.}
    \label{fig:device_2}
\end{figure}

\noindent A wireless capsule endoscope~\cite{fontana2016innovative}, shown in Fig.~\ref{fig:device_2} and further referred to as device~2 (D2), is intended for ColoRectal Cancer (CRC) screening with locomotion control. The primary modules of D2 include an image sensor with optics, an illumination board, a control unit, a telemetry board, an actuation system, a localization unit, and a battery with a recharging circuit. D2 is used in a clinical laboratory and swallowed by the patient. The image sensor of D2 captures the condition of the patient's digestive tract. The collected images are streamed via Bluetooth to the external controller.

\begin{table}[h!]
\renewcommand{\arraystretch}{1.15}
\centering
\caption{Risk matrix - Device 2}
\label{tab:risk_matrix_d2}
\begin{tabular}{|
>{\columncolor[HTML]{E1E8F5}}c |c|c|c|}
\hline
\cellcolor[HTML]{E1E8F5} & \multicolumn{3}{c|}{\cellcolor[HTML]{E1E8F5}\textbf{Impact (\textit{I})}} \\ \cline{2-4} 
\multirow{-2}{*}{\cellcolor[HTML]{E1E8F5}\textbf{Probability (\textit{P})}} & \cellcolor[HTML]{E1E8F5}\textbf{Low} & \cellcolor[HTML]{E1E8F5}\textbf{Moderate} & \cellcolor[HTML]{E1E8F5}\textbf{High} \\ \hline
\textbf{Low} & \cellcolor[HTML]{00B0F0} -- & \cellcolor[HTML]{F8FF00} -- & -- \cellcolor[HTML]{F8A102} \\ \hline
\textbf{Moderate} & \cellcolor[HTML]{32CB00} T5, T7, T8 & \cellcolor[HTML]{F8FF00} -- & \cellcolor[HTML]{F8A102}\begin{tabular}[c]{@{}c@{}} T6 \end{tabular} \\ \hline
\textbf{High} & \cellcolor[HTML]{F8FF00} -- & \cellcolor[HTML]{F8A102}\begin{tabular}[c]{@{}c@{}} -- \end{tabular} & \cellcolor[HTML]{FD6864} -- \\ \hline
\end{tabular}
\end{table}

For D2, the risk matrix is presented in Table~\ref{tab:risk_matrix_d2}. It demonstrates that, among four detected threats, T6 has a high risk and T5, T7, and T8 have low risks.

\begin{figure}[h!]
    \centering
    \includegraphics[width=\columnwidth]{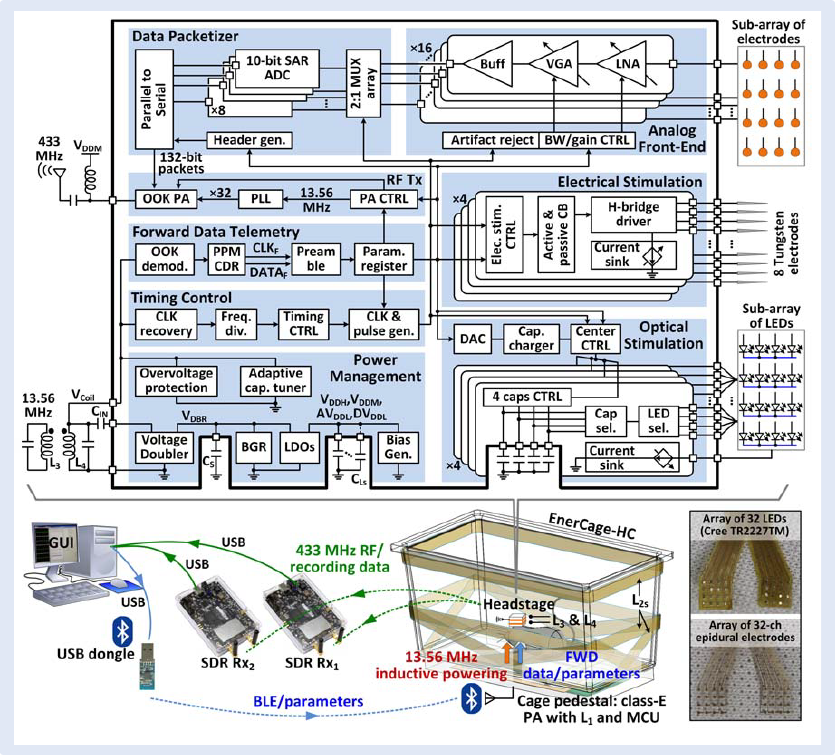}
    \captionsetup{justification=raggedright, singlelinecheck = false}
    \caption{Device 3: block diagram. Adapted from~\cite{jia2020soc}.}
    \label{fig:device_3}
\end{figure}

\subsection{Case study 3 - Implantable}

\noindent A trimodal wireless implantable neural interface SoC~\cite{jia2020soc}, shown in Fig.~\ref{fig:device_3} and further referred to as device~3 (D3), provides optical/electrical stimulation capabilities and neural recordings. D3 consists of optical and electrical stimulation blocks, an analog front-end, a data packetizer, telemetry blocks, a timing control unit, and a power management module. D3 is used in a clinical laboratory. D3 is implanted into the brain and wirelessly paired with a control arena. The recording/stimulation parameters are sent to the arena via BLE from an external terminal (computer). The arena relays the BLE parameters to D3 by on-off-keying (OOK) of a 13.56~MHz power carrier via inductive coils. D3 generates optical/electrical stimulation pulses based on the received parameters. The evoked neural activities are sensed, processed, and transmitted by D3 to the terminal by OOK at 433~MHz. The terminal receives the data by a pair of software-defined radios (SDRs). D3 is not currently used in humans but for scientific experiments in freely behaving animals (rodents). However, it was selected as a case study since its functionality is similar to that of commercial products like~\cite{medtronic2020percept}. In addition, the commercial products themselves are proprietary, which makes them unavailable for the analysis based on public data.

\begin{table}[h!]
\renewcommand{\arraystretch}{1.15}
\centering
\caption{Risk matrix - Device 3}
\label{tab:risk_matrix_d3}
\begin{tabular}{|
>{\columncolor[HTML]{E1E8F5}}c |c|c|c|}
\hline
\cellcolor[HTML]{E1E8F5} & \multicolumn{3}{c|}{\cellcolor[HTML]{E1E8F5}\textbf{Impact (\textit{I})}} \\ \cline{2-4} 
\multirow{-2}{*}{\cellcolor[HTML]{E1E8F5}\textbf{Probability (\textit{P})}} & \cellcolor[HTML]{E1E8F5}\textbf{Low} & \cellcolor[HTML]{E1E8F5}\textbf{Moderate} & \cellcolor[HTML]{E1E8F5}\textbf{High} \\ \hline
\textbf{Low} & \cellcolor[HTML]{00B0F0} -- & \cellcolor[HTML]{F8FF00} -- & -- \cellcolor[HTML]{F8A102} \\ \hline
\textbf{Moderate} & \cellcolor[HTML]{32CB00} -- & \cellcolor[HTML]{F8FF00} -- & \cellcolor[HTML]{F8A102}\begin{tabular}[c]{@{}c@{}} T2, T3 \end{tabular} \\ \hline
\textbf{High} & \cellcolor[HTML]{F8FF00} T4 & \cellcolor[HTML]{F8A102}\begin{tabular}[c]{@{}c@{}} -- \end{tabular} & \cellcolor[HTML]{FD6864} -- \\ \hline
\end{tabular}
\end{table}

For D3, the risk matrix is presented in Table~\ref{tab:risk_matrix_d3}. It shows that T2 and T3 have high risks, whereas T4 has a moderate risk.

\subsection{Case study 4 - Wearable}
\begin{figure}[h!]
    \centering
    \includegraphics[width=\columnwidth]{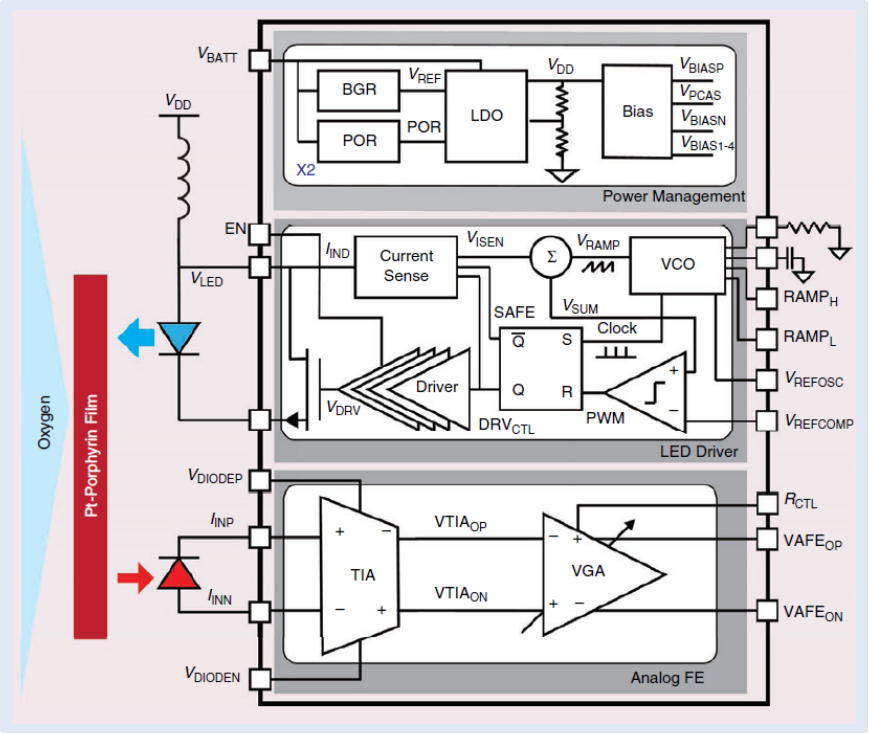}
    \captionsetup{justification=raggedright, singlelinecheck = false}
    \caption{Device 4: block diagram of the readout IC. Adapted from~\cite{2020guler}.}
    \label{fig:device_4}
\end{figure}

\noindent An integrated readout circuit~\cite{costanzo2020integrated}, shown in Fig.~\ref{fig:device_4} and further referred to as device~4 (D4), is used for non-invasive transcutaneous oxygen sensing, which correlates with the blood oxygen level. D4 primary blocks include an analog front-end, a light-emitting diode (LED) driver, and a power management block. D4 is routinely used both in a home setting and in a clinical environment. D4 is intended to be used with dry electrodes in the form of a smart watch or a smart patch. D4 is projected to be a battery-powered device transmitting the sensor data to the external controller via BLE.

\begin{table}[h!]
\renewcommand{\arraystretch}{1.15}
\centering
\caption{Risk matrix - Device 4}
\label{tab:risk_matrix_d4}
\begin{tabular}{|
>{\columncolor[HTML]{E1E8F5}}c |c|c|c|}
\hline
\cellcolor[HTML]{E1E8F5} & \multicolumn{3}{c|}{\cellcolor[HTML]{E1E8F5}\textbf{Impact (\textit{I})}} \\ \cline{2-4} 
\multirow{-2}{*}{\cellcolor[HTML]{E1E8F5}\textbf{Probability (\textit{P})}} & \cellcolor[HTML]{E1E8F5}\textbf{Low} & \cellcolor[HTML]{E1E8F5}\textbf{Moderate} & \cellcolor[HTML]{E1E8F5}\textbf{High} \\ \hline
\textbf{Low} & \cellcolor[HTML]{00B0F0} -- & \cellcolor[HTML]{F8FF00} -- & -- \cellcolor[HTML]{F8A102} \\ \hline
\textbf{Moderate} & \cellcolor[HTML]{32CB00} T5 & \cellcolor[HTML]{F8FF00} -- & \cellcolor[HTML]{F8A102}\begin{tabular}[c]{@{}c@{}} T6, T8 \end{tabular} \\ \hline
\textbf{High} & \cellcolor[HTML]{F8FF00} -- & \cellcolor[HTML]{F8A102}\begin{tabular}[c]{@{}c@{}} -- \end{tabular} & \cellcolor[HTML]{FD6864} -- \\ \hline
\end{tabular}
\end{table}

For D4, the risk matrix is presented in Table~\ref{tab:risk_matrix_d4}. It reveals that T6 and T8 have high risks while T5 has a low risk.              
\section{Discussion and Future Work}
\label{sec:discussion}
\noindent The use of the proposed model revealed and prioritized threats for the case studies of injectables, ingestibles, implantables, and wearables, showing that the model is applicable for a wide range of devices. However, the proposed threat model enables performing further validation, which would involve additional case studies. This validation may be done by investigating known vulnerabilities in devices and comparing the results of the analysis with the outcome of other models. It would also be of interest to include more commercial devices. However, being proprietary and closed source, these devices are challenging to be analyzed based on the public domain information~\cite{rushanan2014sok}. Based on the results of these additional investigations, it may become apparent if separate threat models for each category of MWBDs may provide more information for designers and manufacturers. Another suggestion for future work is to design threat models focused on other primary stakeholders of MWBDs, including manufacturers and hospitals. 
\section{Conclusion}
\label{sec:conclusion}
\noindent This work discussed the importance of security for the emerging miniaturized wireless biomedical devices. The combination of valuable assets belonging to different stakeholders and multiple attack surfaces makes MWBDs a target for cybercriminals. Since MWBDs pose significant risks for their stakeholders, security should be embedded into MWBDs in a structured and repeatable way during the pre-market phase.
The initial step in embedding security into a design is to perform threat modeling. However, it has been shown that MWBDs are distinct from conventional IT devices and require a unique threat model. Therefore, first, this work described the threat modeling process for MWBDs. Then, a domain-specific qualitative-quantitative threat model, suitable for a wide range of MWBDs, was proposed. Among various stakeholders, this model focused on the user.

The model suggested to use six relevant characteristics of attacks to assess their probability and impact. For each characteristic, a three-tiered qualitative-quantitative scale was assigned. The total risk of an attack was defined using the risk matrix approach. The threat model was then applied to the representative designs from each main category of MWBDs. The primary intent of case studies was to detect several threats to show how the model may be adopted by a threat modeling team. The outcomes of the risk analysis reveal that the analyzed devices are vulnerable. The model was easy to apply and sufficient to describe all the observed threats.

\ifCLASSOPTIONcaptionsoff
  \newpage
\fi

\def\UrlBreaks{\do\/\do-}               
\bibliographystyle{IEEEtran} 
\balance                                
\bibliography{references}

\end{document}